\documentclass{article}

\usepackage{preprint_format,times}
\usepackage[utf8]{inputenc}
\usepackage[T1]{fontenc}
\usepackage{microtype}
\usepackage{url}
\usepackage{graphicx}
\usepackage{booktabs}
\usepackage{amsmath,amssymb,amsthm}

\newtheorem{theorem}{Theorem}

\newtheorem{corollary}[theorem]{Corollary}
\theoremstyle{definition}
\newtheorem{definition}[theorem]{Definition}

\newcommand{\Sbar}{\overline S}
\newcommand{\Kbar}{\overline K_c}

\DeclareMathOperator{\sign}{sign}

\title{Abundance Is an Expenditure Condition:\\
AI Automation, Scarcity Rents, and Access}

\author{Song Zichen\\
City University of Hong Kong\\
\texttt{72610558@cityu-dg.edu.cn}}

\preprintfinalcopy

\begin{document}

\maketitle

\begin{abstract}
If AI and automation make reproducible goods cheap, do scarcity rents disappear or migrate? We give a closed general-equilibrium answer that separates a low unit price from economic abundance. Agents consume an automated good, a fixed-supply good, and a numeraire; their purchasing power comes from resource endowments and ownership of scarce assets. The equilibrium of a heterogeneous economy exists, is unique, and is globally stable. Aggregate scarcity rent obeys an exact accounting identity: it rises with automation if and only if real expenditure on the automated good falls. In a symmetric economy this sign is determined by whether the curvature of marginal utility is above or below one. Thus marginal cost converging to zero is insufficient: demand rebound can absorb the released resources. After automated demand is fixed or sufficiently satiated, each agent's access share is an exact convex combination of net purchasing-power and scarce-asset ownership shares. This yields a concentration decomposition with a covariance term, showing why equalizing one asset can increase access inequality when ownership sources were offsetting. A compute extension distinguishes a fixed productive input, whose rent can vanish, from a binding throughput constraint, whose rent persists. Numerical experiments reproduce the sign reversals, ownership effects, and institutional frontiers, and imply eight falsifiable restrictions. The results replace a binary ``post-scarcity'' claim with a testable condition on expenditure, capacity, and ownership.
\end{abstract}

\section{Introduction}

AI systems can lower the marginal cost of software, inference, design, and automated physical production while leaving land, location, energy connections, human attention, and specialized compute capacity in fixed or slowly adjusting supply. The usual narratives jump from this observation to incompatible conclusions: money becomes irrelevant, rents explode, or little changes. Each can occur. The missing step is a model that closes budgets and factor ownership, distinguishes prices from resource use, and states which demand and capacity conditions select among them.

This paper asks: \emph{when reproducible production becomes arbitrarily productive but non-reproducible resources remain scarce, how do prices, rents, ownership, and access reorganize?} The question is not whether all physical constraints vanish. It is whether productivity growth in one sector releases the numeraire resources that finance competition for fixed goods. We call this mechanism \emph{scarcity reallocation}. It has a sharp empirical implication: the direction of a fixed-good rent is governed by real expenditure in the automated sector, not by its unit price alone.

Our minimal economy contains an automated rival good $x$, a fixed-supply good $s$, and a numeraire $z$. Competitive production satisfies $X\leq AZ^f$, hence the automated-good price is $q=1/A$. Agent $i$ owns numeraire $m_i$ and $E_i$ units of the fixed asset, so wealth is $m_i+pE_i$. This ownership term is essential: without it the model describes demand but not who receives scarcity rents. Preferences are $v_i(x_i)+\beta_i\log s_i+\gamma_i\log z_i$, with a concave, potentially satiating $v_i$.

We make five contributions. First, we reduce heterogeneous equilibrium to a one-dimensional excess-demand equation and prove existence, uniqueness, and global price stability. Second, with common $(\beta,\gamma)$ we derive the exact identity
\begin{equation}
 p\Sbar=\frac{\beta}{\gamma}(M-qX),                         \label{eq:rentidentityintro}
\end{equation}
so automation raises scarcity rent exactly when it lowers real automated expenditure $qX$. In the symmetric economy, the sign depends only on relative curvature $\rho_v=-xv''/v'$: rent rises for $\rho_v>1$, is constant for $\rho_v=1$, and falls for $\rho_v<1$. Third, after automated expenditure is determined, access shares are a convex combination of net purchasing-power and asset-ownership shares. Their correlation is a separate distributional state variable. Fourth, we show that physically fixed compute need not earn permanent rent; a binding throughput constraint, rather than a fixed stock by itself, is what sustains compute rent. Fifth, we compare cash, in-kind entitlements, and non-transferable points, and convert the theory into numerical stress tests and falsifiable restrictions.

The main qualification is equally important. This is an equilibrium and measurement framework for AI-induced resource reallocation, not a learning algorithm. Its relevance to the machine-learning community therefore rests on whether the community values a formal account of the economic constraints surrounding large-scale AI. The model's strongest use is as a generator of measurable hypotheses and allocation mechanisms for compute-intensive AI systems.

\section{Related work and gap}

The existence of competitive equilibrium and the complications introduced by satiation are classical \citep{arrow1954existence,aumann1986values,sato2010satiation}. We do not re-prove a general existence theorem. Our contribution is a tractable mixed-abundance economy in which asset income feeds back into demand for the fixed good, yet the equilibrium price remains unique.

Growth theory separates reproducible ideas from rival inputs \citep{romer1990endogenous}, while unbalanced growth and land-rent research show that productivity can redirect expenditure toward relatively inelastic sectors \citep{baumol1967macroeconomics,alonso1964location,rognlie2015deciphering,knoll2017no}. Work on automation studies labor displacement, new tasks, and distribution \citep{acemoglu2019automation,korinek2019artificial,korinek2024scenarios,acemoglu2024simple}. We instead hold the labor margin in the background and isolate the equilibrium interaction of cheap reproducible output, residual expenditure, and ownership of fixed assets.

Digital-goods pricing and cloud allocation analyze nonconvex costs, capacity, and mechanisms \citep{huang2011pricing,ma2014cloud,babaioff2017era}. Fair allocation and market design provide alternatives to money prices \citep{budish2011combinatorial,bogomolnaia2001new,myerson1981optimal}. The unresolved junction is a single closed model that links: (i) demand rebound under falling marginal cost, (ii) persistent or vanishing compute rents, (iii) access inequality generated by joint ownership, and (iv) institutional comparisons. That junction, rather than a prediction that ``markets disappear,'' is our research gap.

\section{Environment and equilibrium}

There are $N\geq2$ agents. Agent $i$ owns $m_i>0$ units of the numeraire and $E_i\geq0$ units of a fixed asset, with $M=\sum_i m_i$ and $\sum_iE_i=\Sbar>0$. A competitive automated sector transforms numeraire input into a reproducible but rival good:
\begin{equation}
 X\leq AZ^f,\qquad A>0.                                    \label{eq:technology}
\end{equation}
The numeraire price is one. Free entry and constant returns imply $q(A)=p_x=1/A$ and zero profit. Agent $i$ has utility
\begin{equation}
 U_i=v_i(x_i)+\beta_i\log s_i+\gamma_i\log z_i,             \label{eq:utility}
\end{equation}
where $\beta_i,\gamma_i>0$, $v_i'>0$, $v_i''<0$, and $v_i'(x)\to0$. Wealth includes asset rent: $w_i(p)=m_i+pE_i$. The budget constraint is
\begin{equation}
 qx_i+ps_i+z_i\leq m_i+pE_i.                                \label{eq:budget}
\end{equation}
The fixed good is a consumption service whose ownership is transferable; consuming it does not destroy the asset. This interpretation covers land services, privileged locations, reservations, and capacity rights.

\begin{definition}[Competitive equilibrium]
Given $A$, an equilibrium is prices $(q,p)$, consumptions $(x_i,s_i,z_i)_{i=1}^N$, and production $(X,Z^f)$ such that: agents maximize \eqref{eq:utility} subject to \eqref{eq:budget}; firms maximize $qX-Z^f$ subject to \eqref{eq:technology}; and $X=\sum_i x_i$, $\sum_i s_i=\Sbar$, and $Z^f+\sum_i z_i=M$.
\end{definition}

This definition distinguishes three notions often collapsed into ``abundance.'' A low unit price means $q\to0$. High supply elasticity follows from \eqref{eq:technology}. We use a resource-adjusted equilibrium definition:
\begin{definition}[Expenditure abundance]
The automated sector is asymptotically abundant along an equilibrium sequence if $q(A)X(A)\to0$.
\end{definition}
This condition says that satisfying endogenous demand consumes a vanishing share of the numeraire resource. It is stronger than $q\to0$ and can fail under demand rebound.

\paragraph{Demand reduction.}
Let $B_i=\beta_i+\gamma_i$, $\delta_i=\beta_i/B_i$, and define residual expenditure after buying $x$:
\begin{equation}
 r_i(w;q)=w-qx_i(w;q).
\end{equation}
At an interior optimum, $x_i$ is the unique solution of
\begin{equation}
 w=q\left[x_i+\frac{B_i}{v_i'(x_i)}\right],                 \label{eq:xi}
\end{equation}
and
\begin{equation}
 s_i=\delta_i\frac{r_i(w_i(p);q)}p,\qquad
 z_i=(1-\delta_i)r_i(w_i(p);q).                             \label{eq:demands}
\end{equation}
Let $\chi_i=\partial r_i/\partial w\in(0,1]$; the endpoint allows an $x_i=0$ corner. Fixed-good excess demand is
\begin{equation}
 F(p)=\sum_i\delta_i\frac{r_i(m_i+pE_i;q)}p-\Sbar.          \label{eq:excess}
\end{equation}

\section{Equilibrium, rebound, and rent migration}

\begin{theorem}[Heterogeneous equilibrium]\label{thm:unique}
Under the stated assumptions, for every $A>0$ there is a unique competitive equilibrium. At its fixed-good price $p^*$,
\begin{equation}
 F'(p^*)=-\frac{\mathcal D}{p^*}<0,\qquad
 \mathcal D=\Sbar-\sum_i\delta_iE_i\chi_i>0.               \label{eq:rootslope}
\end{equation}
Moreover $\dot p=\kappa F(p)$, $\kappa>0$, is globally asymptotically stable on the positive price domain.
\end{theorem}

The asset-income feedback does not create multiple equilibria here. As $p$ rises, owners become wealthier, but each agent devotes strictly less than the full marginal rent to the fixed good: $\delta_i\chi_i<1$. This makes every clearing point a strict downward crossing. Multiple fixed goods with complementarity, nonconvex status utility, or default can break this result.

The same sufficient statistics determine who moves the price. At equilibrium define $\psi_i=\delta_i\chi_i$ and $\mathcal D$ as in \eqref{eq:rootslope}.
\begin{corollary}[Redistribution and price]\label{cor:transfer}
A balanced numeraire transfer from $h$ to $\ell$ and a balanced transfer of fixed-asset rights have, respectively,
\begin{equation}
 \frac{dp}{d\varepsilon}=\frac{\psi_\ell-\psi_h}{\mathcal D},
 \qquad
 \frac{dp}{d\varepsilon}=p\frac{\psi_\ell-\psi_h}{\mathcal D}. \label{eq:transfermain}
\end{equation}
\end{corollary}
Redistributing toward the poorer agent therefore need not lower the scarcity price. What matters is the recipient's marginal propensity to leave wealth for the $(s,z)$ subproblem, not wealth rank itself. This provides a sharper target for transfer experiments than a generic income elasticity.

Now impose common $(\beta,gamma)$ while retaining heterogeneous $v_i,m_i,E_i$.

\begin{theorem}[Expenditure identity and rebound]\label{thm:identity}
Every interior equilibrium satisfies
\begin{equation}
 p\Sbar=\frac{\beta}{\gamma}(M-qX).                          \label{eq:rentidentity}
\end{equation}
Along any differentiable equilibrium branch with $q=1/A$,
\begin{equation}
 \sign\frac{dp}{dA}=\sign(1-\varepsilon_{X,A}),\qquad
 \varepsilon_{X,A}=\frac{d\log X}{d\log A}.                \label{eq:rebound}
\end{equation}
Hence $q\to0$ does not imply higher scarcity rent. If $X/A\to0$, then $p\to\beta M/(\gamma\Sbar)$; if demand expands at least proportionally, the limit changes or fails.
\end{theorem}

Equation \eqref{eq:rentidentity} is an adding-up result, not a forecasting assumption. The numeraire has two final uses: producing $x$ at cost $qX$ and direct consumption $\sum_i z_i=M-qX$. Log separability fixes fixed-good expenditure relative to $z$. It follows that ``scarcity migration'' is conditional: automation redirects rents only when it releases real resources after endogenous demand responds.

For a transparent primitive condition, specialize to $m_i=m=M/N$, $E_i=\Sbar/N$, and common $v$. Symmetry gives $s_i=\Sbar/N$ and
\begin{equation}
 mA=x+\frac{\gamma}{v'(x)},\qquad
 p=\frac{\beta N}{\gamma\Sbar}\left(m-\frac{x}{A}\right). \label{eq:symmetric}
\end{equation}
Define relative curvature $\rho_v(x)=-xv''(x)/v'(x)$ and $t(x)=\gamma/[xv'(x)]$.

\begin{theorem}[Curvature test]\label{thm:curvature}
In the symmetric equilibrium,
\begin{equation}
 \varepsilon_{x,A}=\frac{1+t(x)}{1+\rho_v(x)t(x)},\qquad
 \sign\frac{dp}{dA}=\sign(\rho_v(x)-1).                    \label{eq:curvature}
\end{equation}
Strong curvature ($\rho_v>1$) produces sub-proportional rebound and rising scarcity rent; unit curvature leaves rent unchanged; weak curvature produces super-proportional rebound and falling rent.
\end{theorem}

This test rejects a universal ``AI raises land and resource rents'' claim. It also identifies a measurable sufficient statistic: the elasticity of automated quantity with respect to capability-adjusted productivity. Hard satiation is only one sufficient case; the main theorem does not require a literal consumption ceiling.

\begin{figure}[t]
  \centering
  \includegraphics[width=\linewidth]{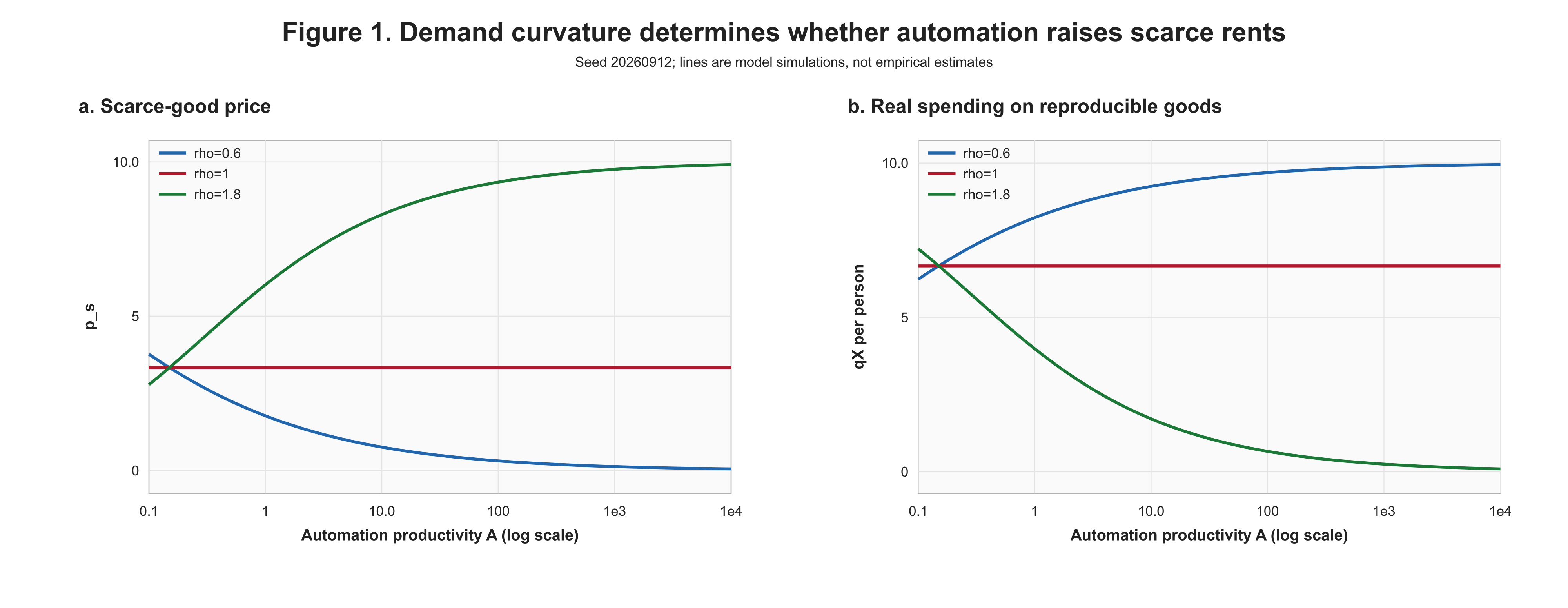}
  \caption{Curvature determines rebound and scarcity rent. We solve \eqref{eq:symmetric} for $v'(x)=2x^{-\rho}$, $m=10$, $\beta=\gamma=1$, and $\Sbar/N=1$. The sign reversal at $\rho=1$ is exact, not fitted.}
  \label{fig:curvature}
\end{figure}

\section{Ownership, access, and compute}

Suppose automated consumption has been fixed by preferences or capacity, and allow the consumer price of $x$ to include compute rent $R_c$. Agent $i$ owns share $\kappa_i$ of compute rent, $\sum_i\kappa_i=1$. Define net purchasing power before fixed-asset rent as
\begin{equation}
 b_i=m_i-p_xx_i+R_c\kappa_i>0,qquad
 Z=\sum_i b_i=M-qX.                                        \label{eq:b}
\end{equation}
With common $(\beta,\gamma)$ let $\delta=\beta/(\beta+\gamma)$, $\mu_i=b_i/Z$, and $\epsilon_i=E_i/\Sbar$.

\begin{theorem}[Access-ownership decomposition]\label{thm:shares}
The unique equilibrium price and individual access shares are
\begin{equation}
 p=\frac{\beta}{\gamma}\frac Z{\Sbar},\qquad
 u_i\equiv\frac{s_i}{\Sbar}=\frac{z_i}{Z}
 =(1-\delta)\mu_i+\delta\epsilon_i.                         \label{eq:shares}
\end{equation}
Let $\widetilde H(a)=\sum_i(a_i-1/N)^2$ and
$\Omega_{\mu,\epsilon}=\sum_i(\mu_i-1/N)(\epsilon_i-1/N)$. Then
\begin{equation}
 \widetilde H(u)=(1-\delta)^2\widetilde H(\mu)
 +\delta^2\widetilde H(\epsilon)
 +2\delta(1-\delta)\Omega_{\mu,\epsilon}.                 \label{eq:hhi}
\end{equation}
The equal-weight planner loss is
\begin{equation}
 W^{SP}-W=(\beta+\gamma)\sum_i\log\frac{1/N}{u_i}\geq0.   \label{eq:welfaregap}
\end{equation}
\end{theorem}

\begin{figure}[t]
  \centering
  \includegraphics[width=\linewidth]{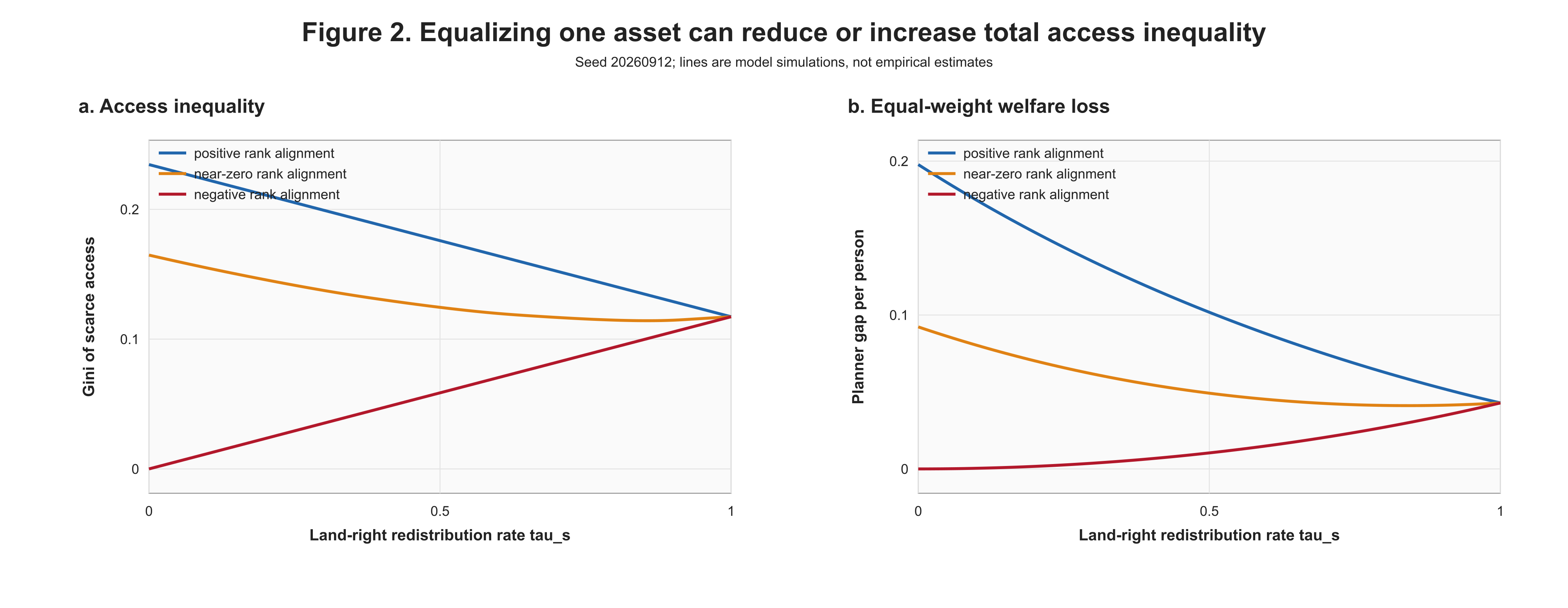}
  \caption{Access depends on joint ownership. With $N=200$ and $\delta=1/2$, complete land equalization reduces access Gini from $0.2345$ to $0.11725$ under positive alignment, but raises it from zero to $0.11725$ under exact negative alignment.}
  \label{fig:ownership}
\end{figure}

The covariance term means the two marginal ownership distributions are insufficient. Equalizing land can raise access inequality when land and purchasing power were negatively aligned and offsetting. Conversely, positive alignment compounds concentration. Figure~\ref{fig:ownership} holds the marginal distributions fixed and changes their rank alignment before progressively equalizing land rights.

\paragraph{Compute is a technology, not a label.}
Consider first a fixed input in a constant-returns technology,
\begin{equation}
 X=A K_c^\eta(Z^f)^{1-\eta},\quad K_c\leq\Kbar,quad0<\eta<1.
\end{equation}
If demand is fixed at $\overline X$ and all $\Kbar$ is used, competitive factor payments imply
\begin{equation}
 Z^f=\left(\frac{\overline X}{A\Kbar^\eta}\right)^{1/(1-\eta)},
 \qquad R_c=\frac{\eta}{1-\eta}Z^f\longrightarrow0.         \label{eq:cdrent}
\end{equation}
A physically fixed stock therefore need not keep a scarcity rent when $A$ raises its effective services.

By contrast, let compute impose throughput $X\leq C(A)$ and $Z^f=qX$. The consumer price is $p_x=q+\nu_c$, where $\nu_c$ is the capacity shadow price. In a symmetric binding allocation $x_c=C/N$,
\begin{equation}
 \nu_c=\frac{v'(x_c)(m-qx_c)}\gamma-q,qquad R_c=\nu_cC.     \label{eq:capacityrent}
\end{equation}
If $x_c\to x_c^\infty\in(0,\infty)$, $qC\to0$, and capacity remains binding, then
\begin{equation}
 R_c\to\frac M\gamma x_c^\infty v'(x_c^\infty)>0.          \label{eq:capacitylimit}
\end{equation}
The long-run loading of compute ownership on access in \eqref{eq:shares} is $(1-\delta)R_c/Z$: it vanishes in \eqref{eq:cdrent} but persists under \eqref{eq:capacitylimit}.

\begin{figure}[t]
  \centering
  \includegraphics[width=\linewidth]{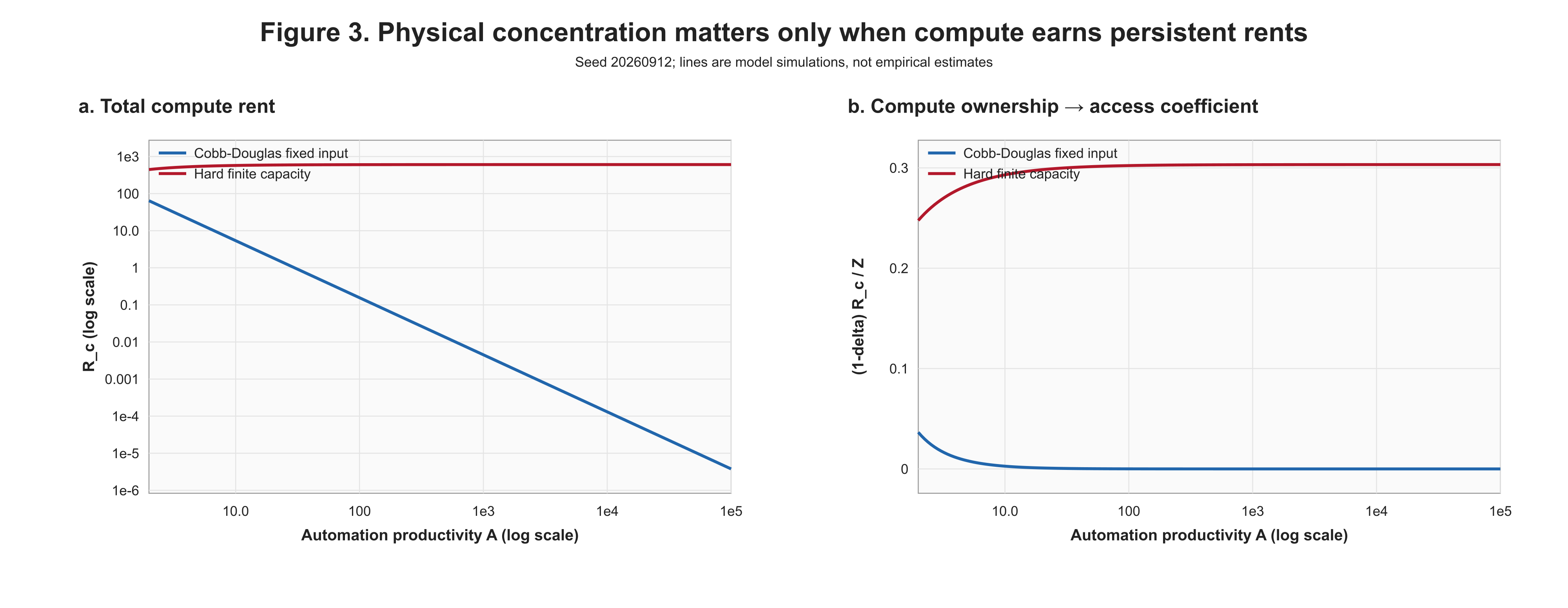}
  \caption{A fixed compute stock and a hard throughput constraint differ. Under fixed-input Cobb--Douglas production, compute rent falls from $64.28$ to $3.78\times10^{-6}$; under a binding finite capacity it rises from $445.50$ to $606.53$. The corresponding terminal ownership transmission coefficients are approximately zero and $0.3033$.}
  \label{fig:compute}
\end{figure}

\section{Institutions and falsifiable implications}

The model separates private welfare, access guarantees, and equal-weight social welfare. No single allocation regime dominates on all three.

\paragraph{Cash versus in-kind abundance.}
Suppose the government provides non-resalable $g_i$ units of $x$ and charges tax $t_i$, with $\sum_i t_i=q\sum_i g_i$. An equal-cost cash policy gives net transfer $qg_i-t_i$. The cash budget set contains the in-kind set, so cash weakly dominates in private utility absent externalities, paternalism, or commitment problems. Yet an access guarantee $x_i\geq g$ requires at least $qNg$ real resources, and direct provision attains this lower bound. Thus in-kind provision has no unconditional welfare advantage, but is cost-minimal for a literal consumption floor.

\paragraph{Markets versus points.}
If the fixed good is public and agents receive non-transferable point budgets $B_i$, clearing yields $s_i=B_i\Sbar/\sum_jB_j$. For heterogeneous fixed-good weights, this allocation is utilitarian-optimal if and only if $B_i\propto\beta_i$. Equal points incur the exact loss $(\sum_i\beta_i)D_{KL}(\theta\Vert\mathbf{1}/N)$, where $\theta_i=\beta_i/\sum_j\beta_j$. Points can equalize access but block mutually beneficial exchange with the numeraire; markets can exploit valuations but encode inherited purchasing power. Their welfare ranking therefore reverses with value-budget alignment and ownership correlation.

\paragraph{Testable restrictions.}
The theory generates restrictions rather than directional slogans: (i) fixed-good rent changes opposite to $qX$; (ii) a balanced transfer from $h$ to $\ell$ changes price with the sign of $\psi_\ell-\psi_h$, where $\psi_i=\delta_i\chi_i$; (iii) post-abundance access loads on purchasing power and ownership with coefficients summing to one; (iv) conditional access concentration rises with $\Omega_{\mu,\epsilon}$; (v) the inequality effect of land equalization interacts with pre-reform rank alignment; and (vi) compute ownership effects scale with $R_c/Z$. Appendix~\ref{app:predictions} gives eight pre-registrable rejection criteria.

These restrictions can be tested in city-industry panels, cloud regions, or controlled allocation experiments. For example, an automation-cost or hardware-compatibility shock can instrument for $A$, while fixed-good prices and real automated-sector inputs identify \eqref{eq:rentidentity}. A randomized transfer design can first estimate $\delta_i$ and the residual-expenditure slope $\chi_i$, then test whether $\psi_i$ predicts the local price response. Joint microdata on cash flow, land or capacity rights, and access can test the adding-up and covariance restrictions in \eqref{eq:shares}--\eqref{eq:hhi}.

\paragraph{Numerical stress tests.}
The experiments are theorem audits, not calibrated forecasts. Each one varies the primitive that a theorem says is decisive while holding the other margins fixed. Table~\ref{tab:numerics} summarizes the endpoint results; the full grids and deterministic code are supplementary.

\begin{table}[t]
\caption{Numerical stress tests. Values are synthetic equilibrium outcomes.}
\label{tab:numerics}
\centering
\small
\begin{tabular}{p{0.24\linewidth}p{0.26\linewidth}p{0.39\linewidth}}
\toprule
Test & Variation & Result \\
\midrule
Demand curvature & $\rho:0.6,1,1.8$ & Scarcity price changes $-3.716,0,+7.131$ as predicted by \eqref{eq:curvature}. \\
Ownership alignment & Same marginals; rank permutation & Full land equalization changes access Gini by $-0.11725$ under positive alignment and $+0.11725$ under exact negative alignment. \\
Compute technology & Fixed input vs. hard capacity & Terminal compute rent is $3.78\times10^{-6}$ versus $606.53$; ownership transmission is $0$ versus $0.3033$. \\
Allocation regime & Cash/in-kind; market/points & An in-kind floor raises minimum $x$ from $2$ to $5$ at loss $0.1672$; market and points welfare cross as ownership alignment varies. \\
\bottomrule
\end{tabular}
\end{table}

\begin{figure}[t]
  \centering
  \includegraphics[width=\linewidth]{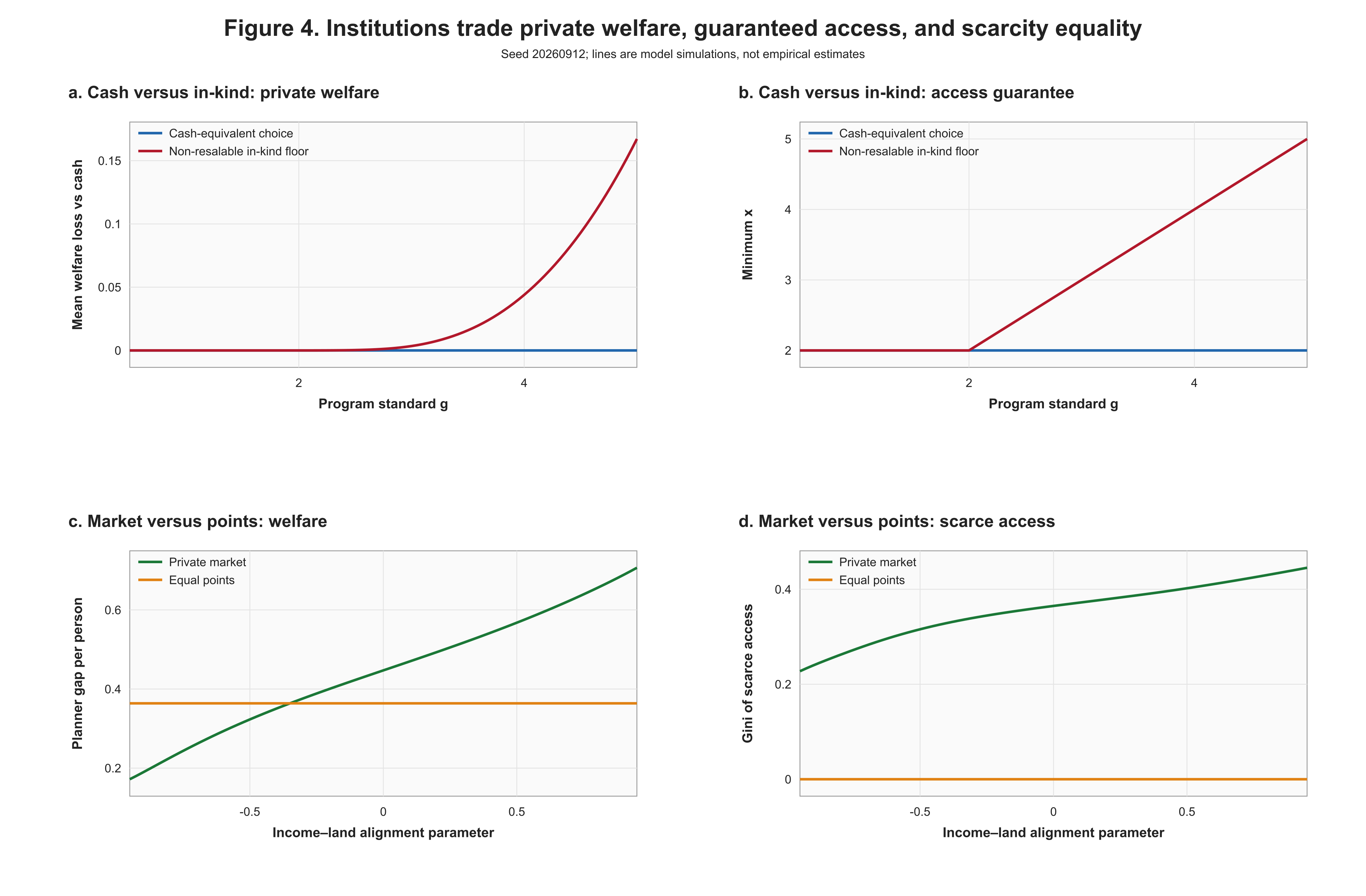}
  \caption{Institutional frontiers. Left: an in-kind floor raises minimum access beyond the voluntary cash allocation only by imposing private-welfare loss. Right: the market-points welfare ranking reverses as purchasing power and ownership become more positively aligned; the simulated crossing is approximately $-0.356$.}
  \label{fig:institutions}
\end{figure}

\section{Discussion and limitations}

Three boundaries discipline the interpretation. First, $x$ is cheaply reproducible but rival; a non-rival digital good requires a public-good or intellectual-property layer. Second, competitive pricing is substantive. Market power can keep $p_x$ above cost even when $q\to0$, and control rights can matter beyond their rental cash flow. Third, the one-fixed-good, log-separable specification buys uniqueness and exact aggregation. Complementary bottlenecks, endogenous construction, migration, credit, political feedback, and positional externalities can generate multiplicity or additional welfare losses.

The model nonetheless clarifies what survives these caveats. Falling unit cost is not sufficient for abundance; a resource or capacity constraint must become slack relative to endogenous demand. Money remains useful because it prices residual scarcity and transfers claims on rent. High absolute consumption can coexist with concentrated access because the latter is pinned to the joint distribution of purchasing power and ownership. Finally, non-monetary allocation changes the ownership of claims and the admissible trades, not the physical scarcity itself.

\section{Conclusion}

AI-driven abundance is an equilibrium expenditure condition, not a property inferred from marginal cost alone. In the baseline economy, scarcity rents rise only when automation releases real resources after demand rebound. Distribution then depends on a convex combination of income and scarce-asset ownership, including their correlation. Compute earns persistent rent only when effective throughput remains binding. These results identify what must be measured before making claims about a post-scarcity economy: real automated expenditure, capacity slack, rental income, and joint ownership. They also turn institutional debate into a transparent frontier among private choice, guaranteed access, and distributional objectives.

\section*{AI Use Statement}

Generative AI tools were used extensively in this research workflow to assist with literature organization, model exploration, algebraic derivations, counterexample construction, simulation code, figure generation, and manuscript drafting. No empirical data were generated or represented as observed data; all reported numerical results are deterministic synthetic simulations of the stated model. AI-assisted mathematical claims were checked against the explicit equilibrium conditions and numerical stress tests. The human authors must complete a final manual verification of all proofs, citations, code, and disclosures before submission and take full responsibility for the paper.

\section*{Ethics Statement}

This paper uses no human-subject or private data. Its allocation results can inform policies affecting access to housing, compute, and public services, but the welfare criteria are deliberately plural and do not prescribe a single regime. Applying the model without measuring externalities, needs, market power, or historical disadvantage could legitimize harmful allocations; these omitted factors should be treated as constraints on policy use.

\section*{Reproducibility Statement}

All primitives and equilibrium conditions are stated in Sections 3--6, and complete proofs are included below. The supplementary package contains a deterministic Python script, parameter grids, generated CSV files, and all plotted figures. No external dataset or proprietary software is required for the numerical results.

\bibliography{references}
\bibliographystyle{preprint_format}

\appendix

\section{Proofs}\label{app:proofs}

\subsection{Consumer reduction and Theorem~\ref{thm:unique}}

The first-order conditions for an interior optimum are
\begin{equation}
 v_i'(x_i)=\lambda_iq,qquad
 \frac{\beta_i}{s_i}=\lambda_ip,qquad
 \frac{\gamma_i}{z_i}=\lambda_i.
\end{equation}
Consequently $ps_i=(\beta_i/\gamma_i)z_i$, and the remaining budget after $x_i$ is divided according to \eqref{eq:demands}. Combining the $x$ and $z$ first-order conditions yields \eqref{eq:xi}. The derivative of the right-hand bracket is
\begin{equation}
 1-B_i\frac{v_i''(x_i)}{[v_i'(x_i)]^2}>1,
\end{equation}
so $x_i$ is unique. Implicit differentiation gives
\begin{equation}
 \chi_i=\frac{\partial r_i}{\partial w}
 =1-\frac{1}{1-B_iv_i''(x_i)/[v_i'(x_i)]^2}\in(0,1).
\end{equation}
At an $x_i=0$ corner, $r_i=w_i$ locally and $\chi_i=1$.

For existence, $r_i(m_i;q)>0$ implies $F(p)\to+\infty$ as $p\downarrow0$. Since $r_i(w;q)\leq w$,
\begin{equation}
 \limsup_{p\to\infty}\sum_i s_i(p)
 \leq\sum_i\delta_iE_i<\sum_iE_i=\Sbar,
\end{equation}
so $F$ is eventually negative. Continuity gives a root. At any smooth root,
\begin{align}
 F'(p)&=\frac{1}{p^2}\left[p\sum_i\delta_iE_i\chi_i-
 \sum_i\delta_ir_i\right]\\
 &=-\frac{1}{p}\left[\Sbar-\sum_i\delta_iE_i\chi_i\right]<0.
\end{align}
The strict inequality follows from $\delta_i\chi_i<1$ and $\sum_iE_i=\Sbar$. One-sided derivatives satisfy the same inequality at a corner. If two roots existed, continuity would require an upward crossing or tangency between two downward crossings, a contradiction. Boundary signs and uniqueness imply $F(p)>0$ below $p^*$ and $F(p)<0$ above it; therefore $\dot p=\kappa F(p)$ converges monotonically to $p^*$. Finally, summing budgets and using $\sum_i s_i=\sum_iE_i$ gives $q\sum_i x_i+\sum_i z_i=M$, which closes production and the numeraire market.

\subsection{Theorems~\ref{thm:identity} and \ref{thm:curvature}}

With common $(\beta,\gamma)$, the first-order conditions imply $ps_i=(\beta/\gamma)z_i$. Summing, clearing $s$, and using numeraire feasibility yields
\begin{equation}
 p\Sbar=\frac\beta\gamma\sum_i z_i
 =\frac\beta\gamma(M-qX),
\end{equation}
which proves \eqref{eq:rentidentity}. Since $q=1/A$,
\begin{equation}
 \frac{d}{dA}\left(\frac XA\right)
 =\frac{X}{A^2}(\varepsilon_{X,A}-1),
\end{equation}
which proves \eqref{eq:rebound}.

In the symmetric economy, market clearing gives $s_i=\Sbar/N$ and the first-order conditions reduce to \eqref{eq:symmetric}. Let $H(x)=x+\gamma/v'(x)=mA$. Then
\begin{equation}
 \varepsilon_{x,A}=\frac{H(x)}{xH'(x)}
 =\frac{1+\gamma/[xv'(x)]}
 {1-\gamma v''(x)/[v'(x)]^2}
 =\frac{1+t}{1+\rho_vt}.
\end{equation}
Since $t>0$, its position relative to one is determined by $\rho_v-1$. Apply Theorem~\ref{thm:identity} to obtain the price sign.

\subsection{Theorem~\ref{thm:shares}}

After $x_i$ is determined, the remaining budget is $ps_i+z_i\leq b_i+pE_i$. The Cobb--Douglas subproblem implies
\begin{equation}
 s_i=\delta\left(\frac{b_i}{p}+E_i\right),\qquad
 z_i=(1-\delta)(b_i+pE_i).
\end{equation}
Summing the first equation and imposing $\sum_i s_i=\Sbar$ gives
\begin{equation}
 p=\frac{\delta}{1-\delta}\frac Z{\Sbar}=\frac\beta\gamma\frac Z{\Sbar}.
\end{equation}
Substitution gives \eqref{eq:shares}. Expanding the square of
$u_i-1/N=(1-\delta)(\mu_i-1/N)+\delta(\epsilon_i-1/N)$ proves \eqref{eq:hhi}. Because $s_i=\Sbar u_i$ and $z_i=Zu_i$, variable utilitarian welfare is
\begin{equation}
 N\beta\log\Sbar+N\gamma\log Z+(\beta+\gamma)\sum_i\log u_i.
\end{equation}
The equal allocation $u_i=1/N$ maximizes the last term by Jensen's inequality, proving \eqref{eq:welfaregap}.

\subsection{Compute-rent derivations}

For the Cobb--Douglas technology, competition and Euler's theorem imply
$Z^f=(1-\eta)p_xX$ and $R_c=\eta p_xX$. Solving the production constraint at $X=\overline X$ yields \eqref{eq:cdrent}; both $Z^f$ and $R_c$ are $O(A^{-1/(1-\eta)})$. For hard capacity, complementary slackness gives $p_x=q+\nu_c$. At the binding symmetric allocation, $x_i=x_c$ and $z_i=m-qx_c$. Combining $v'(x_i)=\lambda_i p_x$ with $\gamma/z_i=\lambda_i$ proves \eqref{eq:capacityrent}; taking the stated limit gives \eqref{eq:capacitylimit}.

\section{Redistribution comparative statics}\label{app:redistribution}

At a clearing price define $\psi_i=\delta_i\chi_i$ and $\mathcal D$ as in \eqref{eq:rootslope}. A balanced numeraire transfer $dm_\ell=d\varepsilon$, $dm_h=-d\varepsilon$ changes the equilibrium price by
\begin{equation}
 \frac{dp}{d\varepsilon}=\frac{\psi_\ell-\psi_h}{\mathcal D}. \label{eq:cashtransfer}
\end{equation}
A balanced transfer of asset rights $dE_\ell=d\varepsilon$, $dE_h=-d\varepsilon$ changes it by
\begin{equation}
 \frac{dp}{d\varepsilon}=p\frac{\psi_\ell-\psi_h}{\mathcal D}. \label{eq:assettransfer}
\end{equation}
Both follow by differentiating $F(p)=0$. The recipient's wealth changes by one unit under the first transfer and by $p$ units under the second. These formulas show why redistribution toward a poorer agent need not lower the scarcity price: the direction depends on the marginal propensity to devote wealth to the $(s,z)$ subproblem, not on wealth rank alone.

\section{Institutional results}\label{app:institutions}

\subsection{Cash and in-kind provision}

Let the government provide non-resalable $g_i$ units of $x$ and levy $t_i$, with $\sum_i t_i=q\sum_i g_i$. Writing additional private purchases as $y_i\geq0$, the in-kind budget is
\begin{equation}
 qy_i+ps_i+z_i\leq m_i+pE_i-t_i,qquad x_i=g_i+y_i.
\end{equation}
Equivalently,
\begin{equation}
 qx_i+ps_i+z_i\leq m_i+pE_i-t_i+qg_i,qquad x_i\geq g_i.
\end{equation}
The equal-cost cash policy has the same budget without $x_i\geq g_i$; its choice set therefore contains the in-kind set. On the other hand, any allocation guaranteeing $x_i\geq g$ for all $i$ uses at least $qNg$ units of the numeraire in production. Direct provision of exactly $g$ attains this physical lower bound. Cash need not attain the target because recipients may optimally spend it on $s$ or $z$.

\subsection{Non-transferable points}

If agents receive point budgets $B_i>0$ and face point price $\pi$, strict monotonicity implies $\pi s_i=B_i$. Clearing gives
\begin{equation}
 \pi=\frac{\sum_jB_j}{\Sbar},\qquad
 s_i=\frac{B_i}{\sum_jB_j}\Sbar.
\end{equation}
For objective $\sum_i\beta_i\log s_i$, the planner assigns $s_i=\beta_i\Sbar/\sum_j\beta_j$. Hence points implement the planner allocation exactly if and only if $B_i\propto\beta_i$. Under equal points, let $\theta_i=\beta_i/\sum_j\beta_j$; the loss is
\begin{equation}
 \sum_i\beta_i\log\frac{\theta_i\Sbar}{\Sbar/N}
 =\left(\sum_i\beta_i\right)D_{KL}(\theta\Vert\mathbf 1/N)\geq0.
\end{equation}
If $\beta_i$ is private information and points increase with reported need, truthful implementation requires an additional incentive mechanism.

\section{Numerical design and additional figure}\label{app:numerics}

All simulations use deterministic grids. The curvature experiment solves \eqref{eq:symmetric} by bisection for $A\in[0.1,10^4]$ and $v'(x)=2x^{-\rho}$, with $\rho\in\{0.6,1,1.8\}$. The ownership experiment uses $N=200$, $\delta=1/2$, fixed marginal distributions of $\mu$ and $\epsilon$, and three permutations controlling rank alignment. The compute experiment compares \eqref{eq:cdrent} with a finite binding-capacity economy at $M=1000$, $\overline X=100$, $\Kbar=10$, and $\eta=0.35$. The institutional experiment draws 500 heterogeneous agents from a fixed pseudorandom seed and compares equal-cost cash with non-resalable in-kind floors, then market shares with equal points across a latent ownership-correlation grid.

\section{Falsifiable predictions and rejection criteria}\label{app:predictions}

\begin{center}
\small
\begin{tabular}{p{0.05\linewidth}p{0.41\linewidth}p{0.44\linewidth}}
\toprule
ID & Prediction & Rejection criterion \\
\midrule
P1 & Fixed-good rent moves opposite to real automated expenditure $qX$. & After controlling total resources and fixed supply, rent rises when $qX$ rises, or falls when $qX$ falls, systematically. \\
P2 & A balanced transfer changes price with $\sign(\psi_\ell-\psi_h)$. & Independently estimated $\psi_i=\delta_i\chi_i$ fails to predict the sign of local price changes. \\
P3 & Post-abundance access loads on net purchasing-power and asset shares with coefficients $(1-\delta,\delta)$. & The coefficient restrictions, including adding to one, are rejected. \\
P4 & Conditional access HHI rises with the ownership-overlap term $\Omega$. & Changing rank alignment while fixing both marginal distributions leaves access HHI unchanged. \\
P5 & Land equalization reduces inequality under positive alignment but can raise it under strong negative alignment. & Reform effects do not interact with pre-reform rank alignment in the predicted direction. \\
P6 & Compute ownership effects scale with $R_c/Z$. & Low-rent and high-rent compute markets exhibit the same ownership transmission after comparable controls. \\
P7 & Equal-cost cash weakly raises private utility; in-kind provision weakly raises target-attainment. & Without externalities or commitment channels, in-kind strictly raises utility without changing target access. \\
P8 & Value-budget mismatch weakens auctions relative to lotteries in total valuation. & Under strong budget constraints, auction assignment follows valuation but not payment capacity. \\
\bottomrule
\end{tabular}
\end{center}

\section{Assumption audit and failure cases}\label{app:audit}

\paragraph{Competition.} Free entry is needed for $q=1/A$. A markup or intellectual-property wedge can keep the consumer price positive even when resource cost vanishes. The expenditure identity then requires separating cost, profit, and factor-rent income.

\paragraph{Log separability.} It generates the exact rent and share identities. More general homothetic $(s,z)$ preferences preserve a related expenditure-share argument; non-homotheticity makes shares wealth-dependent. Complementarity across multiple fixed goods can destroy one-dimensional uniqueness.

\paragraph{Endogenous fixed supply.} If higher rent induces construction, migration, or capacity investment, $\Sbar$ becomes a supply function and price changes combine expenditure and supply responses. The short-run identity remains an accounting benchmark, but the comparative static does not.

\paragraph{Non-rival output.} A non-rival model needs access rights, fixed development cost, and possibly monopoly or public provision. Treating inference or software as rival is appropriate only when serving an additional user consumes capacity or energy.

\paragraph{Status and externalities.} Relative utility can sustain competition even when direct utility is satiated and can create welfare losses absent here. Our welfare results therefore understate the case for corrective policy when positional externalities are strong.

\paragraph{Control rights.} The compute-ownership channel in \eqref{eq:shares} captures cash-flow rights. Ownership may confer agenda-setting power, exclusion, data access, or political influence even when $R_c=0$; those channels require a game rather than a competitive budget model.

\end{document}